\documentclass[%
 aip,
 jmp,
amsmath,amssymb,
reprint,%
]{revtex4-1}

\usepackage{graphicx}

\usepackage{color} 
\usepackage{bm}
\usepackage[utf8]{inputenc}
\usepackage[T1]{fontenc}
\usepackage{mathptmx}
\usepackage{soul} 

\usepackage{float}
\usepackage{hyperref}

\newcommand{\eg}{\textit{e.g.}}

\newcommand{\ie}{\textit{i.e.}}

\begin{document}


\title[Correlation analysis of the dispersion of SARS-CoV-2 in Mexico]
{Correlation analysis of the dispersion of SARS-CoV-2 in Mexico}

\author{M. Flores-Hernandez}
\affiliation{Departamento de Ciencias Naturales y Exactas, Universidad de
 Guadalajara, Carretera Guadalajara - Ameca Km. 45.5 C.P. 46600, Ameca,
 Jalisco, M\'exico.}
\email{pablo.lopez@academicos.udg.mx}

\author{P. C. L\' opez V\' azquez}
\affiliation{Departamento de Ciencias Naturales y Exactas, Universidad de
 Guadalajara, Carretera Guadalajara - Ameca Km. 45.5 C.P. 46600, Ameca,
 Jalisco, M\'exico.}

\author{S. Biswas}
\affiliation{Departamento de Física, Universidad de Guadalajara,
Blvd. Marcelino García Barragan y Calzada Olímpica,
C.P. 44840, Guadalajara, Jalisco, México.}
\email{soham.biswas@academicos.udg.mx}

\date{\today}

\begin{abstract}
In this paper, we propose a method to analyze correlations in pandemic-related data across different 
geographical regions, relying on the analysis of correlations for non-stationary time series, which 
are typical of pandemic data.
Unlike traditional epidemiological approaches focused on medical and modeling perspectives during a 
pandemic, our method emphasizes post-pandemic analysis to assess how societal responses; such as lockdowns, 
travel restrictions, mobility patterns, and vaccination campaigns, manifest in the collective behavior of 
regions. These insights can inform future public health strategies and enhance understanding of the 
complex dynamics underlying pandemic spread and control.
\end{abstract}

\maketitle

\section{\label{intro}Introduction}

Analyzing time series from experimental and real-world data has become a powerful approach for understanding 
complex systems. While stationary cases have been extensively studied; where correlations have proven to be a 
valuable tool, most natural processes are inherently non-stationary. This is particularly true for pandemic 
data, where time series exhibit pronounced non-stationarity due to the complex interplay of biological factors 
(\eg, emerging variants, epidemiological parameters) and societal dynamics (\eg, mobility patterns, confinement 
measures, vaccination campaigns)~\cite{Flaxman2020,Chinazzi20}. Despite this, correlation structures remain 
underexplored in understanding the temporal and spatial evolution of pandemic outbreaks.

In this paper, we introduce a methodology to analyze correlations in pandemic-related data across different 
geographical regions. Unlike traditional epidemiological approaches focused on medical and modeling 
perspectives during a pandemic~\cite{Kermack27,Hethcote2000,Ferguson2020,Pastor15,keeling2008}, our method emphasizes 
post-pandemic analysis to assess how societal responses; such as lockdowns, travel restrictions, mobility patterns, 
and vaccination campaigns, manifest in the collective behavior of regions. Traditional approaches have proven 
instrumental in understanding disease transmission mechanisms, including estimating key epidemiological parameters 
(\eg, the basic reproduction number $R_0$, incubation period, transmission rate) and guiding public health responses 
through compartmental models such as SIR and SEIR framework~\cite{Kermack27,keeling2008}. However, 
these frameworks are largely concerned with biological dynamics, with less attention devoted to quantitatively 
characterizing socio-epidemiological development through spatial correlations. Moreover, whereas existing methods 
often rely on predefined connectivity assumptions, our approach extracts such connectivity directly from real-world 
incidence records.

To address the challenges posed by non-stationary time series, we draw on techniques developed in fields where 
such analyses are well established, particularly financial market analysis
~\cite{munnix2012,chakraborti2020,pharasi2020a,pharasi2021}. These methodologies include the 
use of returns to enhance stationarity, construction of non-degenerate correlation matrices through appropriate 
window selection, eigenvalue spectrum analysis with comparison to Wishart and Marchenko-Pastur distributions, and 
clustering techniques such as $k$-means. By adapting these tools to epidemiological data, we can uncover emergent 
relational structures between regions; such as inter-regional influence, coordination, and response coherence, that 
are not captured by conventional modeling approaches.

We demonstrate the utility of this methodology using COVID-19 as a case study. The pandemic, declared by the WHO 
on March 11, 2020~\cite{who2020}, brought unprecedented changes to mobility, social interactions, and institutional 
operations worldwide. The global scale, rapid spread, and abundance of publicly available data; including 
daily counts of confirmed cases, hospitalizations, and deaths~\cite{Dong2020,OWD,CV19Mex}, make COVID-19 an ideal 
testing ground for post-pandemic analysis that can inform future preparedness strategies. Our analysis focuses on 
the 32 states of Mexico, where data collection artifacts, such as weekly periodicities in reporting~\cite{Lopez2022}, 
necessitate careful preprocessing. We apply band-stop filtering to remove these artifacts and compute returns to 
improve stationarity before constructing correlation matrices across overlapping time windows. This approach 
enables us to examine the temporal evolution of inter-state correlations and identify distinct phases in 
collective behavior during the pandemic.

Our paper is organized as follows: In Section~\ref{data}, we describe the data preprocessing steps, 
including artifact removal and returns calculation, followed by the construction of correlation matrices. 
In Section~\ref{clustering}, we present the clustering analysis of the correlation matrices, revealing distinct regimes of inter-state 
correlation that correspond to different pandemic phases. We also analyze the eigenvalue spectra of the correlation matrices 
to assess the presence of genuine collective behavior. Finally, in Section~\ref{conclusions}, we discuss the implications of our 
findings for understanding pandemic dynamics and inform future public health strategies.

\section{\label{data}Data preparation and correlation analysis}
The data used in this study consists of daily new confirmed COVID-19 cases reported for each of the 32 states in Mexico, 
spanning from February 27, 2020, to December 31, 2022. The data was sourced from the official repository maintained by 
the Mexican government \cite{CV19Mex}, which compiles case counts based on laboratory-confirmed diagnoses and following 
the sentinel surveillance system which consisted of a strategically selected network of reporting units (hospitals, clinics, 
labs) to detect trends, monitor the virus's behavior, and estimate the overall magnitude of an outbreak within 
a population\cite{Re25,Tapia2001}.
Initial inspection of the raw time series revealed pronounced weekly periodicities, likely artifacts of reporting practices
rather than true epidemiological dynamics; this feature in the time series was firstly reported in~\cite{Lopez2022}. 
To mitigate the influence of these artifacts, we applied a band-stop (notch) filter~\cite{smith2007dsp,pei1995notch}
targeting the dominant frequencies identified in the power spectrum of the time series (the amplitude spectrum).
We applied a filter to remove peaks in the power spectrum at periods of one week and its harmonics 
(1/2- and 1/3-week), which were consistently observed across all states. The notch filter 
was implemented to frequencies at the power spectrum corresponding to 
$f_1 = 0.14299706\sim 1/7[1/\mathrm{days}]$, $f_2 =0.28599412\sim 1/3.5[1/\mathrm{days}]$ and 
$f_3 = 0.42899119\sim 1/2.33[1/\mathrm{days}]$ with bandwidths of $\Delta f_1 = (1/7.6- 1/6.35)[1/\mathrm{days}]$, 
$\Delta f_2 = (1/3.7 - 1/3.29)[1/\mathrm{days}]$, and $\Delta f_3 = (1/2.371 - 1/2.3)[1/\mathrm{days}]$. These bandwidths were chosen 
to effectively attenuate the peaks while preserving the relevant frequencies of the signal, ensuring 
that the filtering process did not introduce significant distortions to the underlying epidemiological dynamics.

\begin{figure*}[ht]
\centering
\includegraphics[width=\textwidth]{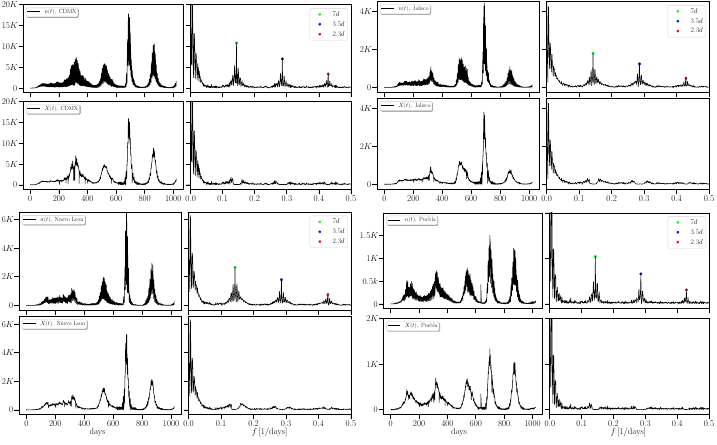}
\caption{Peak filtration: The figure displays the effect of the notch filter on the time series on the states of 
Ciudad de México (top left panels),  Jalisco (top right panels), Nuevo León (bottom left panels), 
and Puebla (bottom right panels). For each state, the first row shows the raw daily new case counts and 
its power spectrum, while the second row shows the filtered time series and its corresponding power spectrum. 
In all cases, the first column depicts the new daily cases (top) and the filtered time series (bottom), 
while the second column shows the power spectrum before (top) and after (bottom) filtering.}
\label{fig:BandFilt}
\end{figure*}

In figure \ref{fig:BandFilt}, we illustrate the effect of the band-stop filter on the time series
from the states of Ciudad de México (top panels) and Jalisco (bottom panels).
The left panel shows the raw daily new case counts, and imposed over them the filtered time series; the panels at the right 
shows the corresponding power spectra before and after filtering.

The use of the notch filter was crucial to remove these artificial periodic components, which could otherwise 
introduce spurious correlations in our analysis. The choice of the band-stop filter was motivated by the fact 
that these frequencies were not punctually identified, so we used a suitable bandwidth around each 
peak. This approach allowed us to selectively attenuate these artificial periodic components while 
preserving relevant frequencies of the signal; we think of it as a 
less shadowing to the relevant correlation signatures of the epidemiological dynamics of the disease.

After filtration, we computed the returns (absolute returns) of the time series to enhance stationarity. 
The returns $R(t)$ are calculated as:
\begin{equation}
R(t) = \frac{X(t+1) - X(t)}{X(t)}
\end{equation}
where $X(t)$ is the filtered daily new case count at time $t$. This transformation
measures the relative change in case counts, effectively normalizing the data and reducing non-stationarity 
over short time windows. In this regard, the returns will help to mitigate non-stationarity and allow 
for more reliable correlation analysis across different states.

\begin{figure}[ht]
\centering
\includegraphics[width=\linewidth]{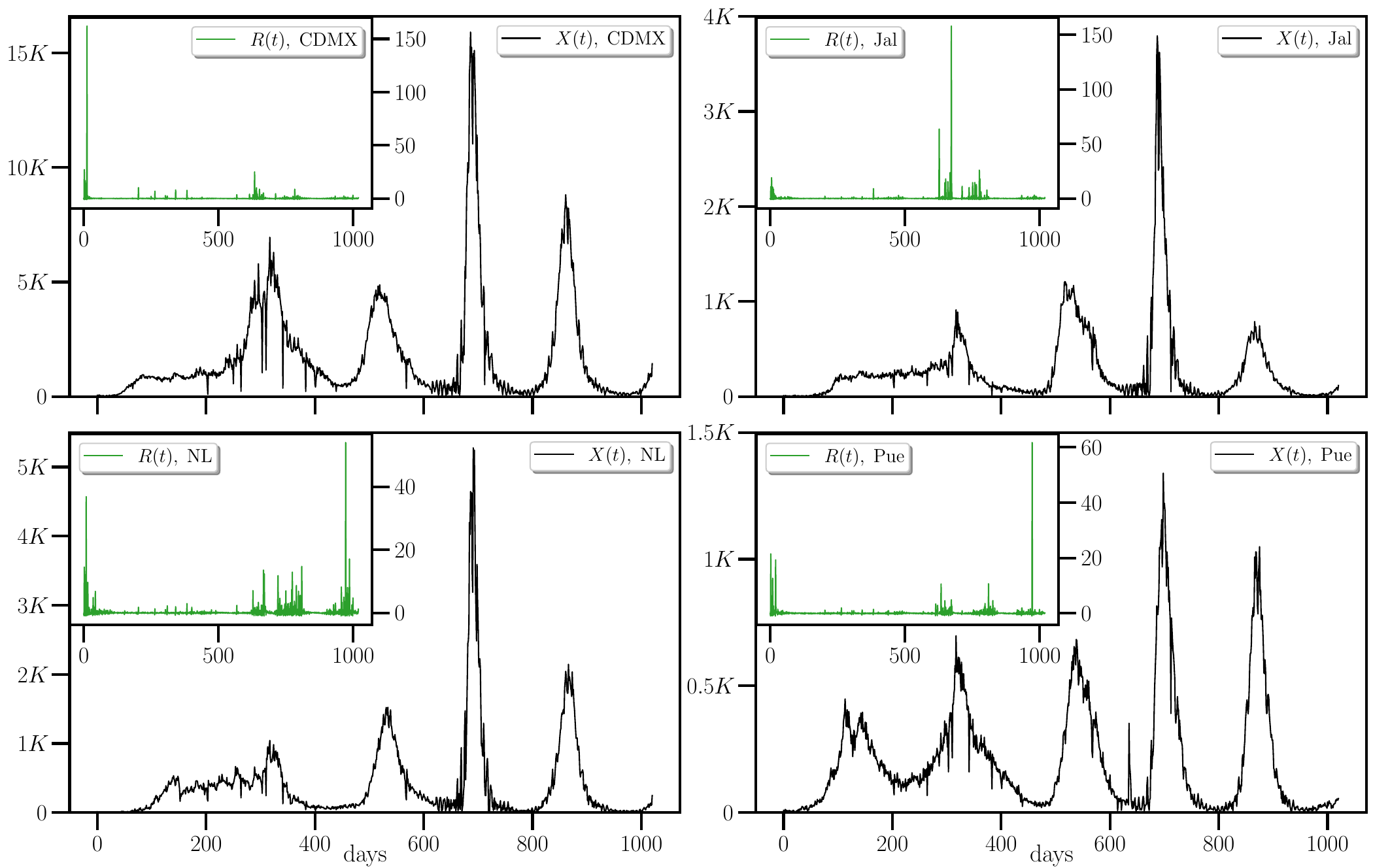}
\caption{Returns time series: The figure displays the returns time series for the states of 
Ciudad de México (top left panels),  Jalisco (top right panels), Nuevo León (bottom left panels), 
and Puebla (bottom right panels). In the figures, the green curves in the inset panels show the returns 
time series of the corresponding filtered data, while the black curves show the original filtered time 
series for comparison.}
\label{fig:Returns}
\end{figure}

Figure \ref{fig:Returns} illustrates the returns time series for Ciudad de México and Jalisco at the top row, and
Sonora and Tabasco at the bottom row, demonstrating the enhanced stationarity (green curves) achieved through
this transformation.

In the figure \ref{fig:Returns}, we can see that the returns time series (green curves) exhibit enhanced 
stationarity compared to the original filtered time series (black curves). In addition, the returns exhibit at certain points 
large peaks, which are due to days with very low or zero new cases followed by days with significant increases in case counts. However, 
despite the presence of large peaks in the returns, these isolated events do not significantly impact the overall correlation structure, 
as they are infrequent and do not dominate the time series.

Now we proceed to compute the correlation matrices for the returns time series. 
Each correlation matrix will capture the pairwise correlations between the returns of the 32 states over a specific 
time window, nevertheless, we have to set an appropriate time window to compute these correlations in order to manage 
the trade-off between having sufficient data points to compute reliable correlations, avoiding degenerate correlation 
matrices, and maintaining temporal resolution to track changes over the course of the pandemic.

In this regard we partitioned the returns time series into overlapping epochs of 33 days, with a 17-day overlap between 
successive epochs.  This particular partition and overlap ensures that the resulting correlation matrices are not degenerate, 
while the overlap preserved sufficient temporal resolution to capture the evolving dynamics of the pandemic. Moreover, robustness 
checks confirmed that the results were not sensitive to moderate variations in epoch length or overlap. This partitioning of 
the 1,021-day dataset yielded 62 epochs per state, from which we computed correlation matrices using the Pearson correlation 
coefficient.

The Pearson correlation coefficient $C_{ij}$ between states $i$ and $j$ is defined as:
\begin{equation}
C_{ij} = \frac{\langle R_i R_j \rangle - \langle R_i \rangle \langle R_j \rangle}{\sigma_i \sigma_j}
\end{equation}
where $R_i$ and $R_j$ are the returns time series for states $i$ and $j$, $\langle \cdot \rangle$ denotes the average over the
epoch, and $\sigma_i$ and $\sigma_j$ are the standard deviations of the returns for states $i$ and $j$ computed within that epoch. 
This coefficient ranges from -1 to 1, with values close to 1 indicating strong positive correlation, values close to -1 indicating strong 
negative correlation, and values near 0 indicating no correlation. By computing $C_{ij}$ for all pairs of states, we obtain a 32x32 
correlation matrix for each epoch, adding up a total of 62 correlation matrices across all epochs corresponding to the entire pandemic.

\begin{figure}[ht]
\centering
\includegraphics[width=\linewidth]{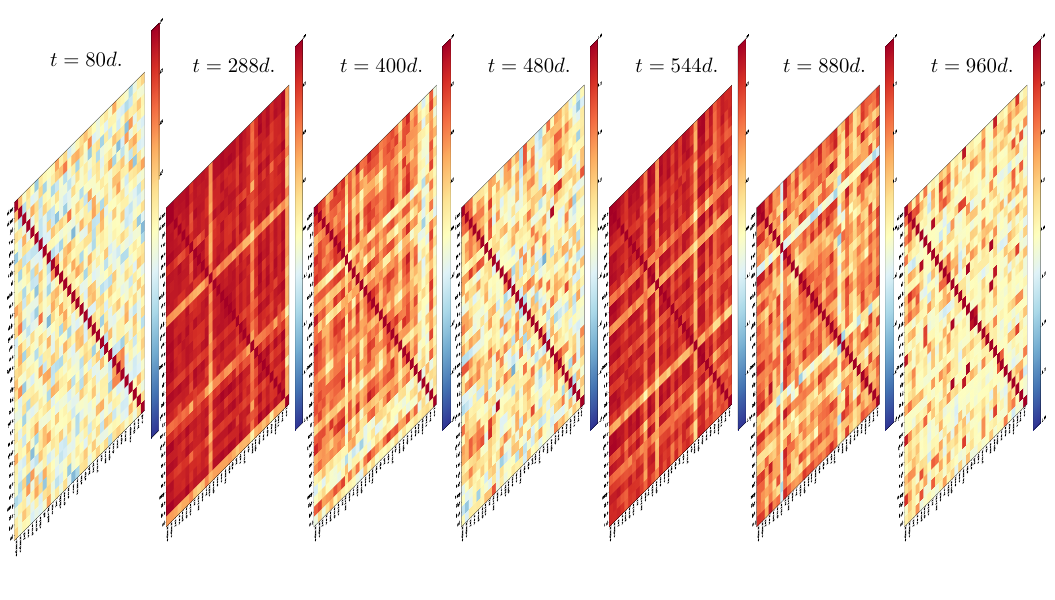}
\caption{Correlation matrices of daily new infections across the 32 Mexican 
states at selected times during the pandemic ($t=80, 288, 400, 480, 544, 880, 960$ days). 
The matrices highlight the temporal evolution of inter-state correlations, 
including periods of strong nationwide synchronization and others characterized by 
block structures reflecting regional heterogeneity. }
\label{fig_CorrMat}
\end{figure}

Figure \ref{fig_CorrMat} illustrates representative correlation matrices at different stages of 
the pandemic. The matrices reveal that correlations among states were not static but evolved 
significantly over time. Certain time windows; such as at $t=80$, $t=480$, and 
$t=960$ days, display similar structures, while other epochs, such as $t=288$ and 
$t=544$, also appear related. These observations suggest that the dynamics of inter-state correlations 
followed recurrent patterns across the pandemic.

In Figure \ref{fig_CorrMat}, two general trends emerge. First, the correlation structure reflects the heterogeneity of the 
pandemic's spread across Mexico: early in the outbreak, correlations were low, indicating 
localized dispersion of the virus, whereas during major waves correlations increased, reflecting widespread 
transmission and more synchronized responses. Second, a block-like clustering of states becomes 
visible during lull periods, suggesting that groups of states experienced similar incidence trends, 
possibly due to shared socio-economic conditions, mobility patterns, or locally coordinated public 
health measures.

The form and structure of the correlation of these matrices suggests that the pandemic dynamics were not uniform across the 
country, but rather exhibited distinct phases of synchronization and regional heterogeneity. These correlation matrices will
form the basis for our subsequent clustering analysis and eigenvalue spectrum examination, allowing us to explore the
temporal evolution of inter-state correlations throughout the pandemic.

\section{\label{clustering}Clustering analysis}

To better characterize the evolving patterns observed in the correlation matrices, we applied a clustering analysis using 
the $k$-means algorithm to the sequence of correlation matrices~\cite{macqueen1967,jain2010}. 
The $k$-means algorithm partitions the set of 62 correlation matrices into $k$ distinct clusters based on feature similarity 
which corresponds to the optimal intra-cluster Euclidean distance. For doing so, one first randomly chooses $k$ initial 
correlation matrices $\tilde{\mathrm{C}}^{(k)}$ (the centroids) from the dataset of correlation matrices, and from which 
the Euclidean distance to all other complementary matrices is computed, \ie:
\begin{equation}
d^{(k)}_l(\mathrm{C}^{(l)}, \tilde{\mathrm{C}}^{(k)}) = \sqrt{\sum_{i,j} (C^{(l)}_{ij} - \tilde{C}_{ij}^{(k)})^2}
\end{equation}
and the correlation matrix $\mathrm{C}^{(l)}$ is assigned to the cluster $k$ 
when the distance $d^{(k)}_l$ to that cluster is minimum, \ie:
\begin{equation}
k^* = \arg\min_k d^{(k)}_l(\mathrm{C}^{(l)}, \tilde{\mathrm{C}}^{(k)})\,.
\end{equation}
Once the assignment of all correlation matrices to the nearest centroid is done, the centroids are updated by calculating 
the mean of the matrices assigned to each cluster, \ie:
\begin{equation}
\tilde{\mathrm{C}}^{(k)} = \frac{1}{|C_k|} \sum_{\mathrm{C}^{(l)} \in C_k} \mathrm{C}^{(l)}\,.
\end{equation}
This process continues until convergence, where the assignments of the new centroids no longer change significantly 
with respect to the previous iteration $(|\tilde{\mathrm{C}}^{(k)} - \tilde{\mathrm{C}}^{(k)}_{\text{prev}}| \leq \epsilon=10^{-6})$, 
or because a predefined number of iterations is reached. Since the initial choice of centroids is random, the algorithm 
consist on finding a global minimum of the intra-cluster distance, which is not guaranteed, so the algorithm is has to be 
run multiple times with different initializations to ensure robustness of the clustering outcome.

The choice of $k$ is crucial and can be guided by methods such as the elbow method, silhouette analysis, or domain 
knowledge~\cite{rousseeuw1987,kaufman1990,tibshirani2001}. 
In our case we tested different values of $k$ and found that four clusters provided the most interpretable and consistent results, 
supported by convergence over $10^5$ algorithmic realizations. The four clusters represent distinct regimes of inter-state correlation 
that correspond to different stages of the pandemic and its management.

\begin{figure}[ht]
\centering
\includegraphics[width=\linewidth]{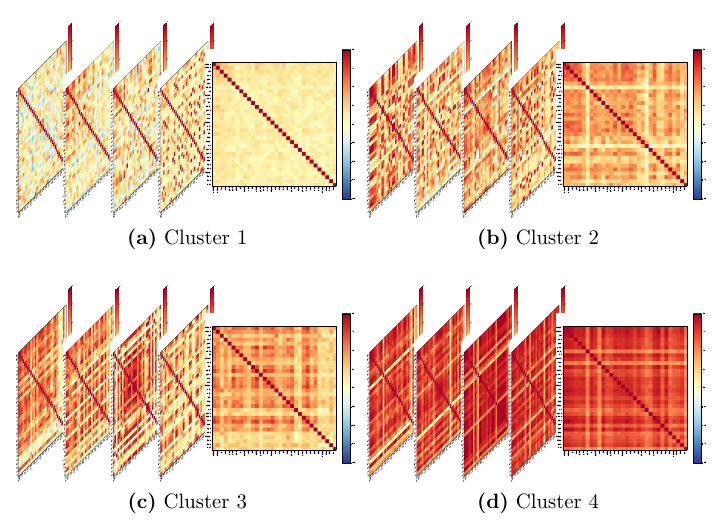}
\caption{Centroids and representative correlation matrices of the four clusters identified by $k$-means analysis.
Panel (a) shows the correlation matrices emerging at times $t= 128, 224, 480, 784$ days (inward perspective), 
together with the centroid of Cluster 1 ($C_1$) (flat perspective) which is characterized by weak correlations, 
with a predominance of near-zero values and some small positive correlations. 
Panel (b) shows the correlation matrices emerging at times $t= 704, 816, 864, 912$ days (inward perspective) 
together with the centroid of Cluster 2 ($C_2$) (flat perspective) which shows intermediate correlation patterns, 
suggesting coordinated but heterogeneous adjustments across states. Panel (c) shows the correlation matrices emerging at 
times $t= 384, 432, 608, 656$ days (inward perspective), together with the centroid of Cluster 3 ($C_3$) (flat perspective) which shows 
intermediate correlation patterns, suggesting coordinated but heterogeneous adjustments across states. 
Panel (d) shows the correlation matrices emerging at times $t= 176, 272, 352, 544$ days (inward perspective), together with 
the centroid of Cluster 4 ($C_4$) (flat perspective) which shows strong, nearly uniform correlations across all states.}
\label{fig:clusters}
\end{figure}

\begin{figure}[ht]
\centering
\includegraphics[width=\linewidth]{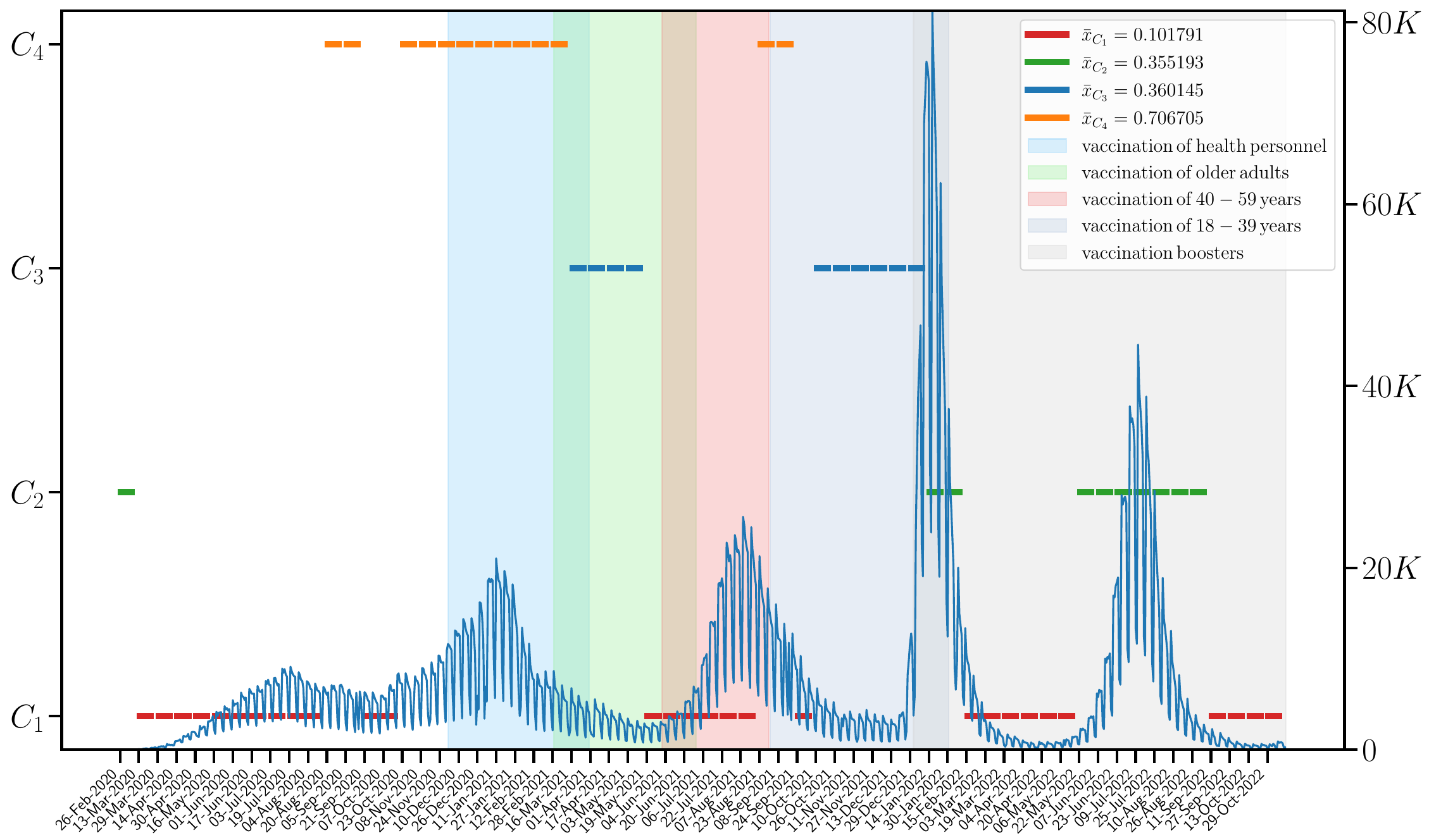}
\caption{Symbolic dynamics of the clusters, aligned with nationwide 
incidence data and major vaccination milestones. The transitions between clusters 
highlight shifts in the collective behavior of states across different pandemic phases. }
\label{fig:SymDyn}
\end{figure}

Figure \ref{fig:clusters} presents the centroids of the four clusters which capture the dominant correlation structure of each cluster, 
together with some representative correlation matrices from each cluster which illustrate how these patterns manifest in specific epochs.
From the figure we observe that Cluster 1 ($C_1$) is characterized by weak correlations, with a predominance of near-zero values and 
some small positive correlations, indicating a regime of localized transmission and heterogeneous state-level responses. Cluster 2 ($C_2$) 
and Cluster 3 ($C_3$) exhibit intermediate correlation patterns showing a more structured correlations that suggest coordinated 
but heterogeneous adjustments across states, likely influenced by vaccination campaigns and the emergence of new viral variants. 
In contrast, Cluster 4 ($C_4$) shows strong, nearly uniform correlations across all states, coinciding with the pandemic peaks
(see Figure \ref{fig:SymDyn}) and high mortality periods—times when nationwide measures and synchronized responses prevailed.  

In Figure \ref{fig:SymDyn} we present the symbolic dynamics of the clusters, which tracks the temporal evolution of the cluster 
assignments for each epoch. The figure reveals how these regimes alternated over time, in close correspondence with pandemic 
phases and major policy interventions. Particularly, the symbolic dynamics is depicted aligned with rough nationwide incidence 
data and major vaccination milestones; in this context, the transitions between clusters highlight shifts in the collective 
behavior of states across different pandemic phases, providing insights into how inter-state correlations evolved in response 
to changing epidemiological conditions and public health interventions. This clustering structure of spatial correlations of 
pandemic data demonstrates the existence of states similar to what has been observed in other complex systems, 
such as financial markets, where distinct correlation regimes have been identified and linked to different market 
conditions~\cite{munnix2012}. The presence of these clusters in pandemic data suggests that similar underlying mechanisms 
of collective behavior and synchronization may be at play.

In Figure \ref{fig:SymDyn}, we notice that the first matrix in the sequence (corresponding 
to the first epoch of the pandemic), belongs to Cluster 2 ($C_2$); we believe this is an artifact of the initial 
conditions of the pandemic where large peaks appeared at the returns time series due to the first cases reported in the country, 
which inturn led to a correlation structure that does not fit neatly into the other clusters. After this initial epoch, 
the dynamics transitions to Cluster 1 ($C_1$), which mostly dominates the early outbreak and inter-wave periods, consistent with 
localized transmission and heterogeneous state-level responses. In contrast, Cluster 4 ($C_4$) shows strong, nearly uniform 
correlations across all states, coinciding with the pandemic peaks and high mortality periods—times before vaccination and 
when strong nationwide measures and synchronized responses prevailed. Cluster 2 ($C_2$) and Cluster 3 ($C_3$) 
exhibit intermediate and more structured correlation patterns, appearing during periods influenced by vaccination 
campaigns and the emergence of new viral variants, suggesting coordinated but heterogeneous adjustments across states. 
Particularly, Cluster 3 ($C_3$) emerges during a later stage of the pandemic, at the end of the healthcare workforce
vaccination and at the beginning of the vaccination of the elder population, coinciding with the periods in between 
the second and the third waves, and the third and the fourth waves, the latter being the most severe wave of the pandemic in Mexico, 
which was driven by the Omicron variant which was characterized by a high transmissibility and a lower severity,. 
The emergence of Cluster 3 ($C_3$) during this period suggests that the vaccination campaign and the emergence of the Omicron variant 
may have led to a more structured correlation pattern among states, reflecting coordinated but heterogeneous adjustments in 
response to these developments. Finally Cluster 2 ($C_2$) emerges after a widespread vaccination of the population, and the 
peaks of the fourth and fifth waves, so the emergence of Cluster 2 ($C_2$) during these periods suggests that the 
widespread vaccination provided relaxation of restrictions and the continued circulation of the Omicron variant. 
The specific correlation matrices belonging to each cluster according to the epoch they belong to, are given in the 
tables \ref{tab:cluster_assignments1} and \ref{tab:cluster_assignments2} at the appendix \ref{app1}.

\begin{figure}[ht]
\centering
\includegraphics[width=\linewidth]{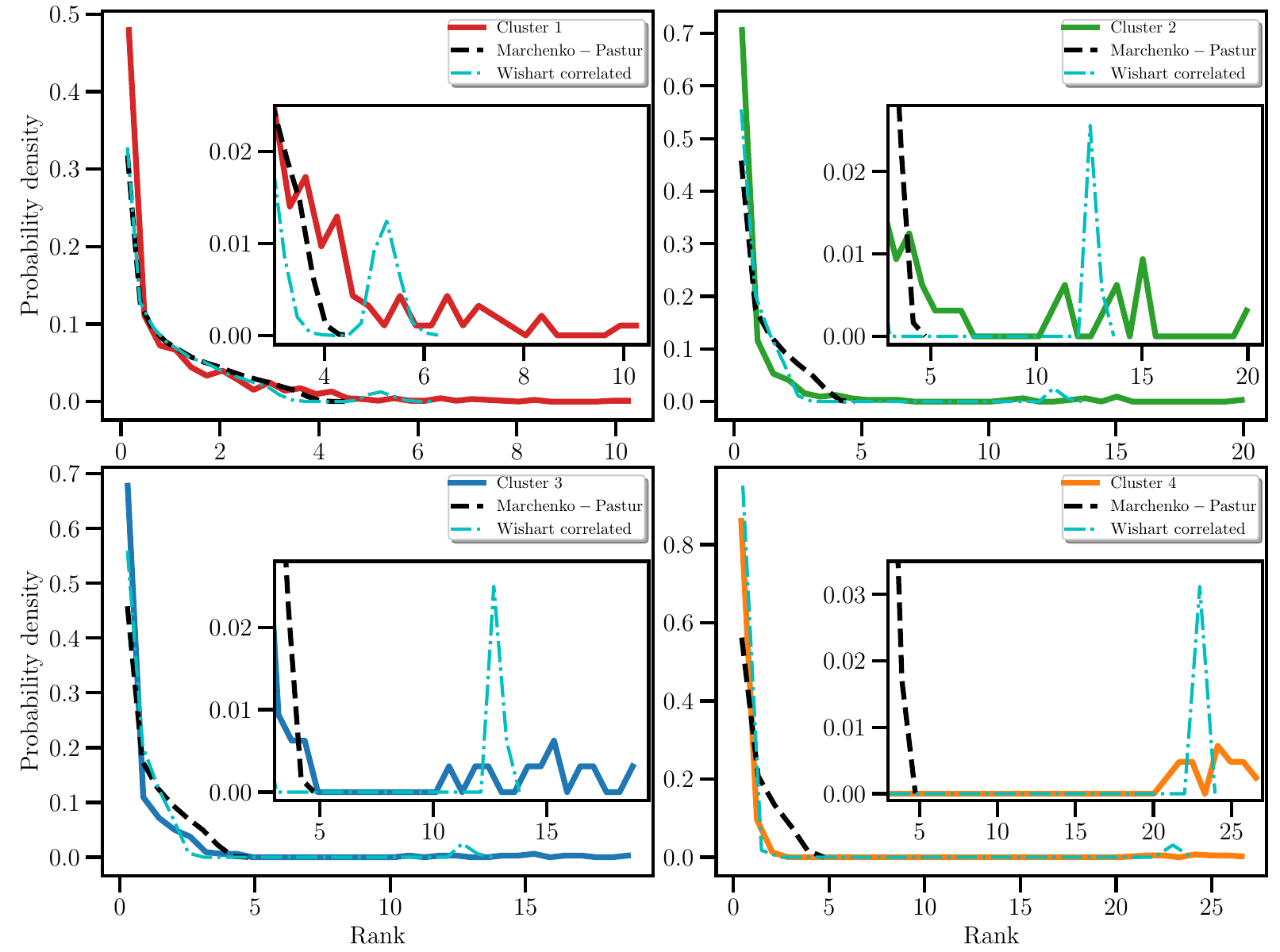}
\caption{Empirical eigenvalue distributions for each cluster compared to theoretical predictions from the Marchenko-Pastur and Wishart 
distributions. The deviations from the theoretical benchmarks indicate the presence of genuine collective behavior and structural 
correlations in the pandemic data.}
\label{fig:Eigen}
\end{figure}

To further validate our assertions, we examined the probability density of the eigenvalues of the correlation matrices in each cluster 
and compare them to the predictions from random matrix theory, \ie, the eigenvalue distributions predicted by the 
Marchenko-Pastur distribution, which describes the eigenvalue spectrum of correlation matrices derived from uncorrelated 
random data~\cite{marchenko1967,mehta2004}, 
and the Wishart distribution, which describes the eigenvalue spectrum of correlation matrices derived from random data 
with constant correlations~\cite{wishart1928,mehta2004}. In the latter, the constant correlations are introduced as the average 
correlation of the empirical 
correlation matrices of each centroid, \ie: $\bar{x}_{C_k} = \frac{1}{N(N-1)} \sum_{i \neq j} \tilde{C}^{(k)}_{ij}$.
By comparing the empirical eigenvalue distributions to these theoretical benchmarks, we can assess the presence of genuine collective 
behavior and structural correlations in the pandemic data, as opposed to random fluctuations. This comparison is presented 
in Figure \ref{fig:Eigen}, where we plot the empirical eigenvalue distributions for each cluster alongside the theoretical predictions from the 
Marchenko-Pastur and Wishart distributions. 
For constructing the probability distributions of the eigenvalues for each cluster, we aggregated the eigenvalues 
from all correlation matrices belonging to that cluster and use a 32-dimensional histogram to estimate the probability density 
distribution of the eigenvalues of the empirical data. In the case of the Wishart distribution and the 
Marchenko-Pastur distribution, we have used a 24 and 17 bins respectively in Cluster 1; 24 and 8 bins respectively in Cluster 2 and Cluster 3;
and 24 and 5 bins respectively in Cluster 4. The number of bins was chosen to provide a better visualization of the distributions, 
while ensuring a sufficient number of eigenvalues per bin to achieve a good balance between resolution and statistical reliability of 
the estimated distributions.

Figure \ref{fig:Eigen} reveals that the eigenvalue distributions of Clusters 1-3 deviate markedly from the 
theoretical benchmarks, particularly through the presence of large eigenvalues, indicating the emergence of genuine collective behavior among states. 
These deviations likely reflect structural correlations driven by mobility, socio-economic ties, and inter-state interactions during 
the pandemic. In contrast, the eigenvalue distribution of Cluster 4 more closely resembles the random benchmarks, suggesting that 
strong confinement measures may have suppressed fine-grained inter-state structures, leading instead to a more homogenized, nationwide response.

\section{\label{conclusions}Conclusions}
In this paper, we introduced a novel methodology for analyzing correlations in pandemic-related data across different geographical regions, 
with a focus on the COVID-19 pandemic in Mexico. By applying techniques from financial market analysis to non-stationary epidemiological 
time series, we were able to uncover distinct regimes of inter-state correlation that correspond to different phases of the pandemic and 
its management. Our clustering analysis revealed periods of localized transmission and heterogeneous responses, as well as times of strong 
nationwide synchronization, particularly during major waves. The eigenvalue spectrum analysis further validated the presence of genuine 
collective behavior among states, driven by mobility and socio-economic interactions. These insights provide a deeper understanding of 
the complex dynamics underlying pandemic spread and control, and can inform future public health strategies. Moreover, our methodology 
could be in principle applied to the analysis of other epidemiological datasets, offering a powerful tool for post-pandemic analysis and 
preparedness planning.

\begin{acknowledgments}
The authors acknowledge financial funding from SECIHTI
through the research project CF 2019/10872. M.F.H. also acknowledges a doctoral fellowship from SECIHTI.
\end{acknowledgments}

\section*{Data Availability Statement}
The data that support the findings of this article are openly
available in the repository of the Mexican government at \url{https://datos.covid-19.conacyt.mx/}.
The full set of correlation matrices together with the processed data 
and the code used for the analysis are available upon reasonable request to the 
corresponding author.

\appendix
\section{\label{app1}Cluster assignments and correlation matrices}
The complete collection of correlation matrices for each epoch, can be found at the supplementary material, while
the specific cluster assignments for each epoch are given in tables \ref{tab:cluster_assignments1} and 
\ref{tab:cluster_assignments2}.
Table is split in two parts for better visualization.
\begin{table}[ht]
\centering
\caption{Cluster assignments for each epoch.  Epochs are numbered from $t=0$ corresponding to the first 
epoch(part 1).}
\begin{tabular}{|c|c|c|c|c|}
\hline
\textbf{Epoch} & \textbf{Cluster 1} & \textbf{Cluster 2} & \textbf{Cluster 3} & \textbf{Cluster 4} \\
\hline
$t=0-15$ days     & & $\mathrm{C}^{(0)}$ & & \\ 
$t=16-31$ days    & $\mathrm{C}^{(1)}$ & & & \\ 
$t=32-47$ days    & $\mathrm{C}^{(2)}$ & & & \\
$t=48-63$ days    & $\mathrm{C}^{(3)}$ & & & \\
$t=64-79$ days    & $\mathrm{C}^{(4)}$ & & & \\
$t=80-95$ days    & $\mathrm{C}^{(5)}$ & & & \\
$t=96-111$ days   & $\mathrm{C}^{(6)}$ & & & \\
$t=112-127$ days  & $\mathrm{C}^{(7)}$ & & & \\
$t=128-143$ days  & $\mathrm{C}^{(8)}$ & & & \\
$t=144-159$ days  & $\mathrm{C}^{(9)}$ & & & \\
$t=160-175$ days & $\mathrm{C}^{(10)}$ & & & \\
$t=176-191$ days & & & & $\mathrm{C}^{(11)}$ \\ 
$t=192-207$ days & & & & $\mathrm{C}^{(12)}$ \\
$t=208-223$ days & & $\mathrm{C}^{(13)}$ & & \\ 
$t=224-239$ days & & $\mathrm{C}^{(14)}$ & & \\
$t=240-255$ days & & & & $\mathrm{C}^{(15)}$  \\ 
$t=256-271$ days & & & & $\mathrm{C}^{(16)}$  \\
$t=272-287$ days & & & & $\mathrm{C}^{(17)}$ \\
$t=288-303$ days & & & & $\mathrm{C}^{(18)}$ \\
$t=304-319$ days & & & & $\mathrm{C}^{(19)}$ \\
$t=320-335$ days & & & & $\mathrm{C}^{(20)}$ \\
$t=336-351$ days & & & & $\mathrm{C}^{(21)}$ \\
$t=352-367$ days & & & & $\mathrm{C}^{(22)}$  \\
$t=368-383$ days & & & & $\mathrm{C}^{(23)}$  \\  
$t=384-399$ days & & & $\mathrm{C}^{(24)}$ & \\ 
$t=400-415$ days & & & $\mathrm{C}^{(25)}$ &  \\
$t=416-431$ days & & & $\mathrm{C}^{(26)}$ &  \\
$t=432-447$ days & & & $\mathrm{C}^{(27)}$ &  \\
$t=448-463$ days & $\mathrm{C}^{(28)}$ & & & \\  
$t=464-479$ days & $\mathrm{C}^{(29)}$ & & & \\
$t=480-495$ days & $\mathrm{C}^{(30)}$ & & & \\
$t=496-511$ days & $\mathrm{C}^{(31)}$ & & & \\
\hline
\end{tabular}
\label{tab:cluster_assignments1}
\end{table}

\begin{table}[ht]
\centering
\caption{Cluster assignments for each epoch.  Epochs are numbered from $t=0$ corresponding to the first 
epoch(part 2).}
\begin{tabular}{|c|c|c|c|c|}
\hline
\textbf{Epoch} & \textbf{Cluster 1} & \textbf{Cluster 2} & \textbf{Cluster 3} & \textbf{Cluster 4} \\
\hline
$t=496-511$ days & $\mathrm{C}^{(31)}$ & & & \\
$t=512-527$ days & $\mathrm{C}^{(32)}$ & & & \\
$t=528-543$ days & $\mathrm{C}^{(33)}$ & & & \\
$t=544-559$ days & & & & $\mathrm{C}^{(34)}$ \\ 
$t=560-575$ days & & & & $\mathrm{C}^{(35)}$ \\
$t=576-591$ days & $\mathrm{C}^{(36)}$ & & & \\ 
$t=592-607$ days & & & $\mathrm{C}^{(37)}$ & \\ 
$t=608-623$ days & & & $\mathrm{C}^{(38)}$ & \\
$t=624-639$ days & & & $\mathrm{C}^{(39)}$ & \\
$t=640-655$ days & & & $\mathrm{C}^{(40)}$ & \\
$t=656-671$ days & & & $\mathrm{C}^{(41)}$ & \\
$t=672-687$ days & & & $\mathrm{C}^{(42)}$ & \\
$t=688-703$ days & & $\mathrm{C}^{(43)}$ & & \\ 
$t=704-719$ days & & $\mathrm{C}^{(44)}$ & &\\
$t=720-735$ days & $\mathrm{C}^{(45)}$ & & & \\ 
$t=736-751$ days & $\mathrm{C}^{(46)}$ & & & \\
$t=752-767$ days & $\mathrm{C}^{(47)}$ & & & \\
$t=768-783$ days & $\mathrm{C}^{(48)}$ & & & \\
$t=784-799$ days & $\mathrm{C}^{(49)}$ & & & \\
$t=800-815$ days & $\mathrm{C}^{(50)}$ & & & \\
$t=816-831$ days & & $\mathrm{C}^{(51)}$ & & \\ 
$t=832-847$ days & & $\mathrm{C}^{(52)}$ & & \\
$t=848-863$ days & & $\mathrm{C}^{(53)}$ & & \\
$t=864-879$ days & & $\mathrm{C}^{(54)}$ & & \\
$t=880-895$ days & & $\mathrm{C}^{(55)}$ & & \\
$t=896-911$ days & & $\mathrm{C}^{(56)}$ & & \\
$t=912-927$ days & & $\mathrm{C}^{(57)}$ & & \\
$t=928-943$ days & $\mathrm{C}^{(58)}$ & & & \\ 
$t=944-959$ days & $\mathrm{C}^{(59)}$ & & & \\
$t=960-975$ days & $\mathrm{C}^{(60)}$ & & & \\
$t=976-991$ days & $\mathrm{C}^{(61)}$ & & & \\
\hline
\end{tabular}
\label{tab:cluster_assignments2}
\end{table}

\bibliography{man}   

\end{document}